\documentstyle[aps,preprint]{revtex}   
\begin{document}
\draft

\title{The effect of a modulated flux on the growth of thin films}

\author{Pablo Jensen$^*$ and Bernd Niemeyer}

\address{ D\'epartement de Physique des Mat\'eriaux, Universit\'e
Claude Bernard Lyon-1, 69622 Villeurbanne C\'edex, France}

\maketitle

\begin{abstract}
Thin films are usually obtained by depositing atoms with a continuous
flux. We show that using a chopped flux changes the growth
and the morphology of the film. A simple scaling analysis predicts how
the island densities change as a function of the frequency of the
chopped flux in simple cases where aggregation is irreversible.
These predictions are confirmed by
computer simulations. We show that the model can be used to obtain 
information on the diffusion or the evaporation of the adatoms. 
The model is also useful to understand the growth of thin films
prepared by pulsed sources. 

\end{abstract}

\narrowtext

One of the main interests of usual deposition techniques such as
Molecular Beam Epitaxy \cite{mbe} is that the structure of the
deposited films is to a large extent determined by kinetic factors, as
opposed to thermodynamic equilibrium. This allows to ``play games''
\cite{lagally} with the different growth parameters (incident flux of
particles, diffusion coefficient of an adatom \ldots) in order to
obtain different film morphologies. A simple example is given by the
quantity of islands grown on a substrate at low enough temperatures :
it is known that the number of islands at saturation is given by
$(F/D)^{1/3}$ \cite{villain,venables,stoyanov} where $F$ is the
incident flux and $D$ the diffusion coefficient. Then, by increasing
the flux or decreasing the diffusion constant (by lowering the
substrate temperature), one can adjust the saturation number of
islands grown on the substrate. In this sense, each kinetic factor is
a ``handle'' on the system, allowing to control the morphology of the
films.  We introduce in this Letter a new kinetic handle, which should
enable a larger control over film growth : the {\it chopping} of the
incident flux. We note that this flux modulation is intrinsic to other
deposition techniques such as cluster laser vaporization (the laser is 
pulsed \cite{perez}). It is therefore important to understand
how growth proceeds in the presence of a modulated flux if one is to be able
to interpret experiments performed in these conditions. For example, one
may wonder whether the usual growth theories  \cite{villain,venables,stoyanov}
can be used by replacing the continuous flux by the average
value of the chopped flux over a cycle. In the following, we will show 
that this is not the case, and that the growth of the film is profoundly 
changed by the modulation of the incident flux for the case of growth with 
{\it irreversible} aggregation (critical island size 1, see 
\cite{villain,venables,stoyanov,ss,brune,bartdimer}. Conversely, we show
what kind of information can be derived from
experiments carried under these conditions.

The basic idea of our method is that if instead of using a continuous
flux we use a {\it chopped} flux to grow a film, the number of islands
formed on a substrate will depend on the chopping frequency $f$ and on
$d$, the fraction of the period the flux is "on" (see Fig. \
\ref{hache}). This dependence is due to the fact that the free
particle concentration on the surface does not reach its steady state
concentration instantaneously, but only after a time which we will
call $\tau_m$. Then, if the timescale of the chopping (1/f) is much
smaller than $\tau_m$, the system only sees the average flux. In the
contrary case, everything happens as if the instantaneous flux was
used instead. Then, there will be a transition from one behaviour to
the other at a chopping frequency close to $1/\tau_m$.  We discuss
this in detail in next section in the case of growth without
desorption: the corresponding analysis in presence of evaporation is
briefly presented afterwards.

\vspace{.5cm} {\it Presentation of the method :} Fig \ \ref{hache}
shows schematically the time evolution of the free adatom (monomer) concentration 
in presence of a modulated flux. We assume here, for simplicity, that 
adatom-adatom aggregation is irreversible and that only monomers can 
move on the
surface \cite{villain,venables,stoyanov,ss,brune,bartdimer}. Then, the
important timescale for growth is given by the time needed for
monomers to reach their steady state concentration or to
disappear. Since, in the absence of evaporation, the monomers
disappear mainly by diffusing randomly until they aggregate with an
island \cite{villain}, their mean lifetime on the surface is given by
$\tau_m \sim l^2/D$ where $2 l$ is the mean distance between islands
 and $D$ the diffusion constant of the monomers. We obtain
$\tau_m \sim 1 /(D N)$ where $N$ is the island concentration.  We
can predict three regimes of behaviour, depending on the relative
magnitude of $\tau_m$, d/f and 1/f.

For low chopping frequencies ($\tau_m \ll d/f$), the monomer
concentration reaches its steady state concentration $\rho_{ss} = F
\tau_m$ almost instantaneously in the timescale of a
period.  After the flux is turned off, the monomer concentration
goes back to 0 also almost instantaneously ($\tau_m \ll 1/f$). 
Then, between two successive "flux on" periods, nothing
happens since only the monomers can move, and there is no monomer
left. Therefore, growth proceeds as if we had a continuous flux $F_i$
and the island concentration at saturation for low frequencies
$N_{lf}$ satisfies the well-known result $N_{lf} \sim (F_i / D)^\chi$,
with $\chi \sim 0.36$ for fractal islands \cite{villain}.

When the frequency increases ($d/f \ll \tau_m$), the monomer
concentration cannot attain its steady state concentration during the
"flux on" part of the cycle : $\rho(t=d/f) \sim F_i d/f \ll
\rho_{ss}$.  Since we are interested in $d \ll 1$, $\rho$ vanishes
before next cycle, and the average of the monomer concentration over a
cycle is smaller than in the preceding regime, leading to a smaller
island density. More precisely, the increase of island density during
a cycle is given by : $\Delta N = \int_{cycle} D \rho^2(t) dt$. Using
$\rho(t) = (F_i d/f) e^{-t/\tau_m}$, we obtain $\Delta N  \sim D
\int_{d/f}^{\infty} (F_i d/f) e^{-2t/\tau_m} \sim 
1/2\left(F_id/f\right)^2 * (D \tau_m)$. Since $\tau_m \sim 1/DN$, we
obtain $\Delta (N^2) \sim \left(F_id/f\right)^2$. We have checked that, as
in the case of a continuous flux \cite{villain,evap}, the saturation
island density is reached for a surface coverage of about 15\%. Then,
the total number of cycles needed to reach
saturation is $n_c \sim f/(6F_id)$. Since $\Delta (N^2)$ is
a constant over the cycles, we simply multiply $\Delta (N^2)$ by the
number of cycles to obtain $N_i$ the saturation island
density in this intermediate regime : $N_i \sim (F_id/f)^{1/2}$. We note
that here the saturation island density does {\it not} depend on the
diffusion constant. This regime is limited by the fact that if $\tau_m
\sim 1/f$, the monomer concentration no longer fully decays when the
next "on" cycle begins. The monomer concentration slowly increases,
reaching its steady state value after several cycles.

We then cross to the regime of high frequencies ($\tau_m \gg 1/f$),
where many deposition cycles are carried out during the monomer
equilibration, and the system only sees the average flux
$F_{av}$. Then the island concentration at saturation for high
frequencies $N_{hf}$ satisfies $N_{hf} \sim (F_{av} / D )^ \chi =
d^\chi N_{lf} \ll N_{lf}$.

\vspace{.5cm}

{\it Measuring the evaporation time :} We briefly treat the case of
film growth in presence of significant evaporation of monomers from
the surface (for a more detailed treatment in the presence of a
continuous flux, see \cite{evap}). In this case, the lifetime of the
adatoms is no longer determined by the aggregation with islands but
rather by random evaporation after a mean time $\tau_e$
\cite{venables,stoyanov,evap}.  We study here the dependence of the
island density $N^{e}$ at a {\it fixed} amount of deposited matter as a
function of the modulation frequency.  This quantity is not
necessarily the saturation island density but it is easier to measure
experimentally when evaporation is important. As in the non
evaporation case, three different regimes will be scanned as the
frequency increases.

For low chopping frequencies ($\tau_e \ll d/f$), the monomer
concentration reaches its steady state concentration $\rho_{ss} = F
\tau_e$ and vanishes almost instantaneously in the timescale of a
period. Therefore, during a cycle, the island density increases
according to $dN^{e}/dt \sim D \rho^2 \sim D F_i^2 \tau_e^2$, leading to
$\Delta N^{e} (per \ cycle) = D F_i^2 \tau_e^2 d / f$. The island density
for fixed $M = F_{av} t$ in this regime is therefore $N^{e}_{lf} \sim D
F_i \tau_e M$ since the number of cycles is $n_c = fM/(d F_i)$.

When the frequency increases ($d/f \ll \tau_e$), the monomer
concentration cannot attain its steady state concentration during the
"flux on" part of the cycle : $\rho(t=d/f) \sim F_i d/f \ll
\rho_{ss}$.  Again, the increase of island density during a cycle is
given by : $\Delta N^{e} = \int_{cycle} D \rho^2(t) dt \sim 1/2
\left(F_id/f\right)^2 D \tau_e$. Counting the number of cycles as
before leads to an island density in this intermediate regime $N^{e}_{i}
\sim D F_i d \tau_e M f^{-1}$. As in the case of complete
condensation, this regime is limited by the fact that if $\tau_e \sim
1/f$, the monomer concentration no longer fully decays when the next
"on" cycle begins. Then, the monomer concentration slowly increases,
reaching its steady state value after several cycles.

In the regime of high frequencies ($\tau_e \gg 1/f$), many deposition
cycles are carried out during the monomer equilibration, and the
system only sees the average flux $F_{av}$. Using the result obtained in 
the low frequency case (see also Ref. \cite{evap}), the island concentration
 at saturation for high frequencies can be written as $N^{e}_{hf} \sim D
F_{av} \tau_e M = d N^{e}_{lf} \ll N^{e}_{lf}$.

\vspace{.5cm}

{\it Computer simulations :} Monte-Carlo simulations of the growth of
thin films by a continuous flux have been independently implemented by
several groups with small variations \cite{bales,boston,model}. To
simulate the growth in the presence of a chopped flux, we introduce
the four following physical processes :

(1) {\it Deposition}.  Particles are deposited at randomly-chosen
positions of the surface at a flux $F_i$ during the ``on'' fraction of
a cycle ($d \ll 1$). During the rest of the cycle, no particle reaches
the surface (see Fig \ \ref{hache}).  The average flux reaching the
surface is thus $F_{av} = d F_i$.

(2) {\it Diffusion}.  Each diffusion time $\tau$, all monomers
 are chosen at random and moved in a random direction by one
lattice site.

{(3)} {\it Evaporation}. Isolated adatoms can evaporate off the
surface at a constant rate. We denote by $\tau_e$ the mean lifetime of
a free adatom on the surface.

(4) {\it Aggregation.} If two particles come to occupy neighboring
sites, they stick irreversibly and form an island. Islands are assumed
to be immobile.

We first analyse the results when desorption is negligible ($\tau_e =
\infty$).  Fig \ \ref{noevap} shows the evolution of the island
density for a coverage of 15\% (which we have checked is also the
maximum island density, see also \cite{villain}) as a function of the
rescaled frequency $f/d$. We see the three regimes predicted by our
simple analysis. In particular, the relation $N_i \sim f^{-1/2}$
is well verified in the intermediate regime. The inset shows that the
ratio of the low and high frequencies island densities scales as
$d^\chi$, as predicted.

When desorption is important, the agreement is also very good.  Fig \
\ref{evap} shows that the relation $N_{i}^{e} \sim f^{-1}$ is well
verified in the intermediate regime. As in the case of complete
condensation, the inset in Fig \ \ref{evap} shows that the ratio of
island densities in the low and high frequencies regimes scales as
$d$.

\vspace{.5cm}

{\it Summary, Perspectives :} As anticipated, we have shown that it
is not possible to interpret growth experiments under a chopped
flux in the framework of usual growth models \cite{villain,venables,stoyanov} 
which assume a continuous flux. For example, the relation of the 
saturation island density
to the experimental parameters (flux, diffusion coefficient) {\it
depends} on the regime, i.e. on the frequency of the chopped flux.

What kind of information can be derived from flux-modulated
experiments for usual deposition techniques such as Molecular Beam Epitaxy?
We note that modulation of the flux
can be achieved rather easily by intercalating a rotating disk with a
slit close to the sample.  The order of magnitude of the frequencies
required to observe the three regimes is reasonable. For example, in
Stroscio's Fe/Fe(001) experiments \cite{stroscio}, at room
temperature, $N \sim 7 \ 10^{12} cm^{-2}$ and $D \sim 7 \ 10^{-12}
cm^{2} s^{-1}$ then $f_{high} \sim 4DN \sim 200 Hz$ and $f_{low} = d
f_h \sim 2 Hz$ for $d=0.01$. In the case of complete condensation, measuring 
the saturation island density $N_{sat}$
as a function of frequency can help checking the growth
conditions. First, $N_{sat}$ should show a sharp decrease for frequencies
close to $1/\tau_m$. Second, the ratio of the densities in the two extreme
regimes should scale as $d^\chi$ if aggregation is indeed irreversible. We
notice that this test - which is similar to the $N_{sat}$ versus flux
test \cite{villain,ss,brune,flux} - can be carried out experimentally
without changing the flux, or even without knowing it. The effects of 
{\it reversible} aggregation should be addressed in future studies.
The intermediate regime is also interesting since $N_{sat}$ is independent
of the diffusion constant (and therefore of temperature). In growth
experiments carried under a continuous flux, this constant density may
only be observed for growth in presence of trapping impurities or 
reconstructed surfaces. In the case of growth in presence of important 
desorption, we speculate that this method allows to measure
the evaporation time {\it in-situ}.  By using reflection-high-energy-electron
diffraction (RHEED) one could in
principle measure a characteristic time needed for growing a monolayer,
say a quarter oscillation. Although RHEED oscillations are difficult
to interpret in detail, we anticipate that this characteristic time
should depend strongly on the modulation frequency in the neighborhood
of $f \simeq 1/\tau_e$. Therefore, studying the evolution of this 
characteristic time as a function of the chopping frequency may lead to 
an estimate of $\tau_e$.

\begin{figure}
\caption{Schematic illustration of the chopped flux (bold) and the
corresponding evolution of the monomer density (dashed line). $F_i$ and
$F_{av}$ refer to the instantaneous and average flux respectively.}
\label{hache}
\end{figure}

\begin{figure}
\caption{Monte Carlo simulation of the island densities as a function of
the rescaled chopping frequency f/d in the case of complete condensation.
The island densities correspond
to a surface coverage of 15\%, an instantaneous flux $F_i = 10^{-7}$
and a diffusion constant $D=1/4$.  Each curve corresponds to a
different value of $d$ : $d=.5$ (circles), $d=.1$ (squares),
$d=.01$ (diamonds), $d=.001$ (triangles). The solid line has a slope of
-1/2. The inset shows the
dependence of the high frequency island density on $d$. The solid line
is a fit to the data and has a slope 0.36, in excellent agreement with the predicted slope
($\chi = 0.36$).}
\label{noevap}
\end{figure}

\begin{figure}
\caption{Computer simulation of the island densities as a function of
the rescaled chopping frequency f/d when growth is limited by evaporation.
The island densities correspond to a deposition time $F_{av} t = 0.77$,
an instantaneous flux $F_i =
10^{-7}$, a diffusion constant $D=1/4$ and an evaporation time $\tau_e = 100$.
 Each curve corresponds to
a different value of $d$ $d=.5$ (left triangles), $d=.1$
(squares), $d=.01$ (circles), $d=.003$ (triangles up), $d=.001$
(diamonds). The solid line has a slope -1. 
The inset shows the dependence of the high frequency
island density on $d$. The solid line is a fit to the data and
has a slope 0.97, in excellent agreement with the predicted slope of 1.}
\label{evap}
\end{figure}

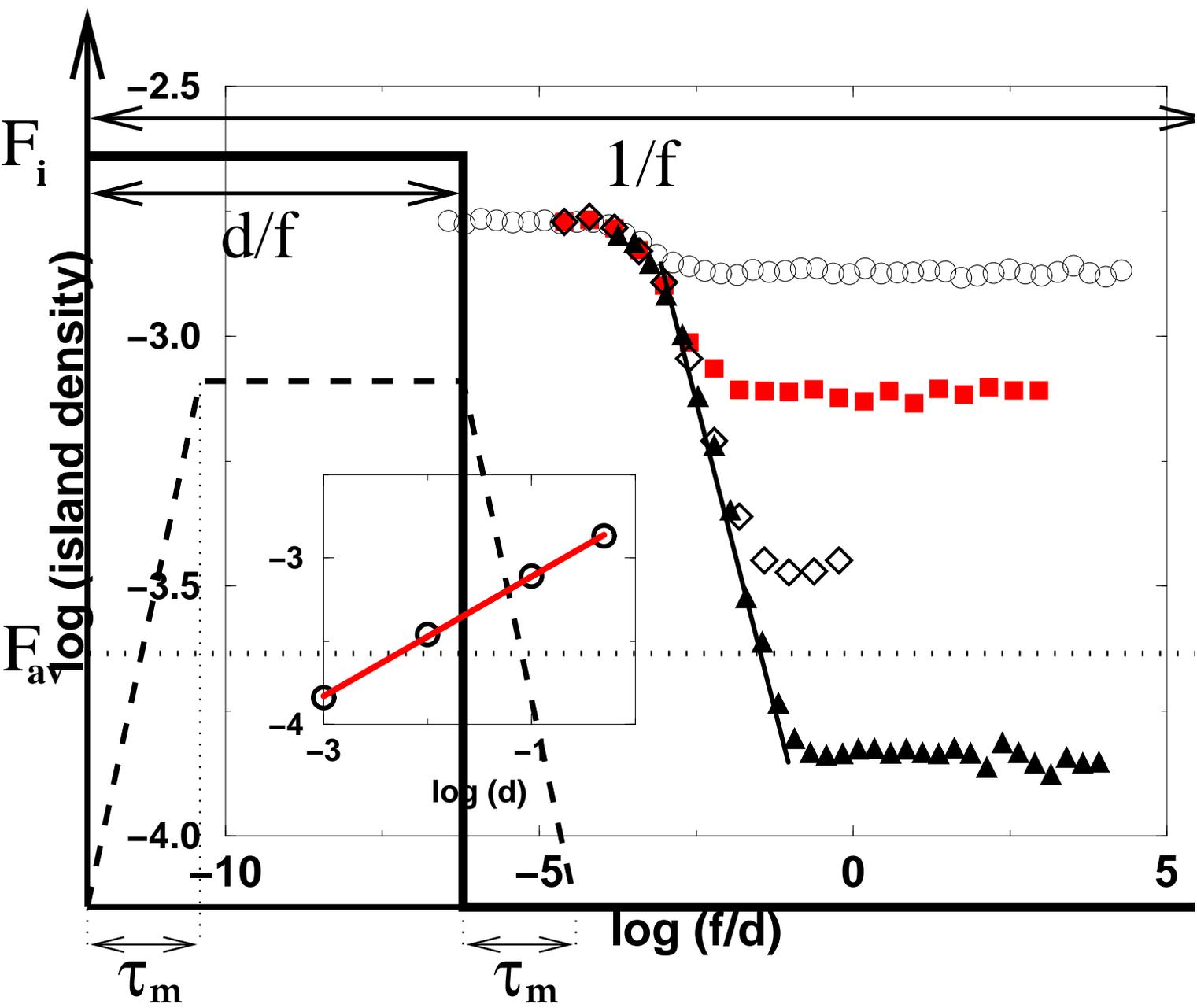
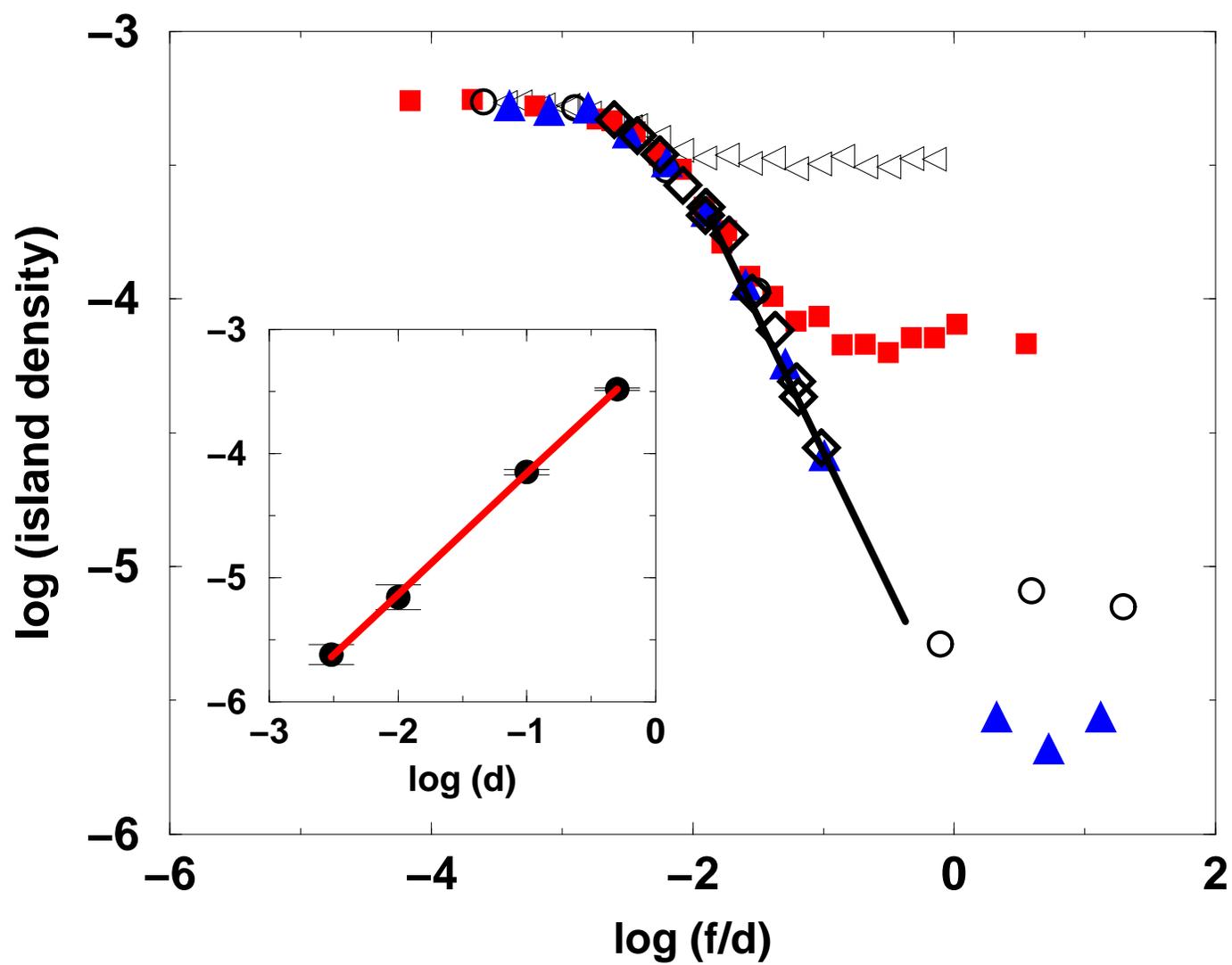
\end{document}